\documentclass[
 preprint,
 nofootinbib,
 amsmath,
 amssymb,
 aps,
 showpacs
]{revtex4-1}

\usepackage{graphicx}
\usepackage{dcolumn}
\usepackage{bm}
\usepackage{citesort}
\usepackage{color}

\newcommand{\fref}[1]{Fig.~\ref{#1}}
\newcommand{\Fref}[1]{Figure \ref{#1}}
\newcommand{\eref}[1]{Eq.~(\ref{#1})}

\begin{document}

\title{Experimental noiseless linear amplification using weak measurements}

\author{Joseph Ho, Allen Boston, Matthew Palsson and Geoff Pryde}

\email{g.pryde@griffith.edu.au}

\affiliation{Centre for Quantum Computation and Communication Technology (CQC2T) \\ and Centre for Quantum Dynamics, Griffith University, Brisbane, 4111, Australia}

\date{\today}

\begin{abstract}

The viability of quantum communication schemes rely on sending quantum states of light over long distances. However, transmission loss can degrade the signal strength, adding noise. 
Heralded noiseless amplification of a quantum signal can provide a solution by enabling longer direct transmission distances and by enabling entanglement distillation. The central idea of heralded noiseless amplification---a conditional modification of the probability distribution over photon number of an optical quantum state---is suggestive of a parallel with weak measurement: in a weak measurement, learning partial information about an observable leads to a conditional back-action of a commensurate size.
Here we experimentally investigate the application of weak, or variable-strength, measurements to the task of heralded amplification, by using a quantum logic gate to weakly couple a small single-optical-mode quantum state (the signal) to an ancilla photon (the meter).
The weak measurement is carried out by choosing the measurement basis of the meter photon and, by conditioning on the meter outcomes, the signal is amplified.
We characterise the gain of the amplifier as a function of the measurement strength, and use interferometric methods to show that the operation preserves the coherence of the signal.

\end{abstract}


\maketitle

\section{Introduction \label{sec:intro}}
Quantum control techniques provide tools for transforming quantum states and thus for handling and processing quantum information. The essence of control approaches is to feed back information---obtained from a measurement, for example---to change the state of the system. In quantum mechanics, the simplest feedback is the inherent back-action of the quantum measurement itself. By conditioning on a subset of measurement outcomes, quantum control can produce transformations beyond the set allowed by unitary evolutions alone~\cite{Nielsen2000}.

A projective, or strong, measurement collapses the state of the system onto an eigenstate of the observable measured. It is possible to vary the strength of the measurement, however, trading off the amount of information obtained for the amount of back-action caused~\cite{Pryde2004,Ralph2006}. These \textit{weak measurements} \footnote{Note that the term ``weak measurements'' is sometimes used to refer to those cases with extremely small back-action; in that notation non-projective measurements might be called general-strength measurements. We use the term weak to mean ``not strong''.} can be modelled or implemented using the von Neumann measurement model, treating the measurement apparatus (meter), like the signal, as a quantum system. The two systems undergo an entangling unitary operation. Subsequently a projective measurement of the meter retrieves outcomes corresponding to some physical property of the signal. However, as the two systems are coupled, this necessarily provides a back-action onto the signal \cite{Ralph2006}. By reducing the interaction strength in the measurement, disturbance of the signal can be reduced at the cost of information gained.

We consider the use of weak measurement back-action to implement the task of noiseless linear amplification (NLA; also called heralded noiseless amplification). 
NLA holds prospect for enabling long-distance quantum communication schemes, for overcoming loss, by amplifying the quantum signal noiselessly, preserving quantum coherences and allowing recovery of the original quantum state's features.
In classical networks, which use macroscopic signals, amplification is readily exploited in devices called repeaters, however deterministically amplifying a quantum signal noiselessly is forbidden by quantum mechanics as it violates the `no-cloning' theorem \cite{Wootters1982}.
Indeed, quantum limits of deterministic amplifiers, which are phase-insensitive (linear), have been well studied and shown to add a minimum amount of noise \cite{Caves1982}.
However, the authors in Ref.~\cite{Ralph2009} have shown, by construction of a non-deterministic operator, that NLA can be implemented probabilistically provided the operation of the amplifier, on average, does not increase the distinguishability of any two non-orthogonal states.
The nondeterminism is addressed by an independent heralding mechanism to sort successful outcomes from the failed cases, which can then be used to implement schemes such as entanglement distillation and probabilistic state cloning \cite{Ralph2009}.
Experimental realisations, using the quantum scissors technique, have demonstrated key features of such an amplifier in the setting of continuous \cite{Xiang2010,Ferreyrol2010} and discrete variables \cite{Kocsis2012,Osorio2012,Bruno2013}.
An alternative method for heralded NLA has been realised by successive application of creation and annihilation operators on a optical mode \cite{Zavatta2010}. Our techniques complement and advance the existing experimental work by realising heralded amplification through a different model.

Our work is related to an existing theory proposal~\cite{Menzies2009} based on the idea of the weak values formalism, which combines weak measurements with pre- and post-selection on input states. We explore the use of weak measurements, but without the use of weak values, to implement heralded NLA of small coherent states.
Our experimental demonstration requires modest resources by using linear optical quantum logic techniques and single photon counting methods. Our work is also closely related to a theory proposal to use generalised controlled-unitary operations to implement NLA functionality~\cite{McMahon2014}.

\section{Theory \label{sec:theory}}
\begin{figure}
\includegraphics{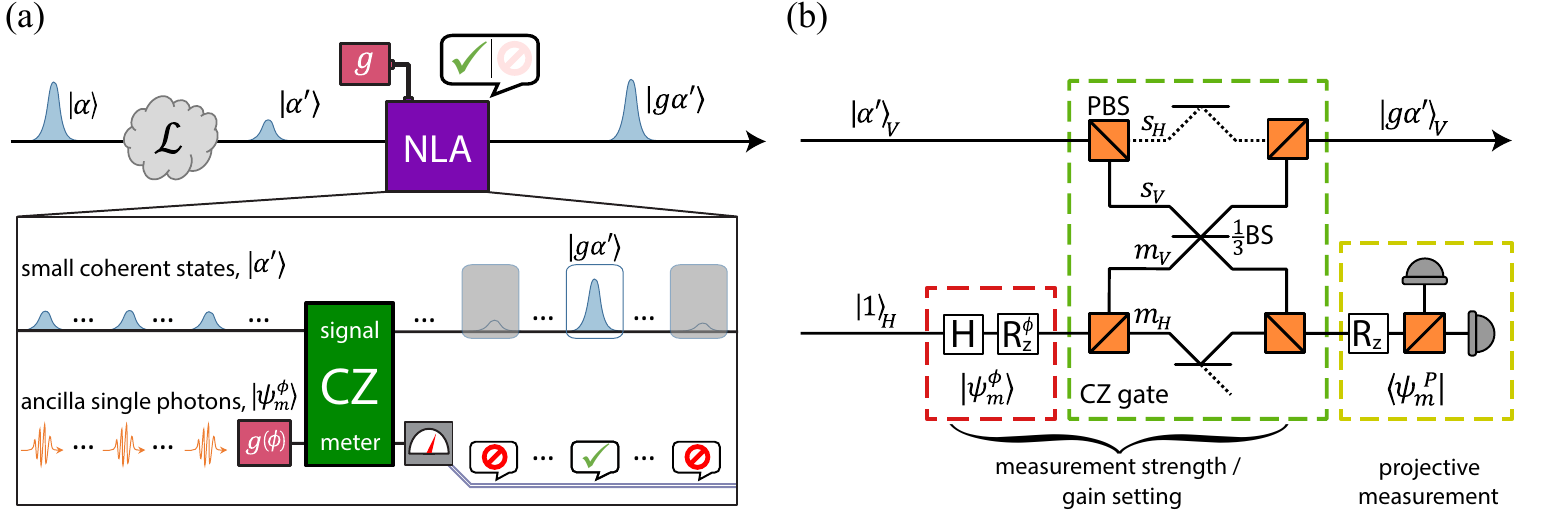}
\caption{\label{fig:concept} (a) 
Transmission loss, $\mathcal{L}$, in an optical channel reduces the size of a quantum state of light: $|\alpha\rangle \rightarrow |\alpha ' \rangle$. 
Heralded NLA can be used to amplify small coherent states $|\alpha'\rangle \rightarrow |g\alpha'\rangle$ with an amplitude gain $g$, without introducing noise. 
Our weak measurement NLA protocol is implemented using a CZ gate to weakly couple the signal mode to an ancilla photon, $|\psi_m^\phi\rangle$.
Conditional on the meter measurement outcomes the signal states can be sorted into cases where the back-action has amplified the state.
(b)
Schematically the small coherent state, $|\alpha'\rangle{_V}\approx\mathcal{N}\left(|0\rangle+\alpha|1\rangle\right)_V$, interacts with the meter qubit in the CZ gate by occupying the vertical mode of the signal, $s_V$ in the linear optics CZ gate \cite{Ralph2002}.
The measurement strength is coherently tuned by varying the meter state parameter $\phi$ which sets the gain of the amplifier.
A projective measurement on the meter induces back-action onto the signal, driving the NLA state transformation conditional on the measurement outcome of the meter.}
\end{figure}

\Fref{fig:concept}(a) illustrates the concept of our scheme for realising heralded NLA using weak measurements.
The signal is encoded as a coherent state of light, which is expressed in Fock state notation as $ |\alpha \rangle = \exp{\left(-|\alpha |^2/2\right)}  \sum_{n} \alpha ^n/\sqrt{n!} |n \rangle{},$ where $n$ is the photon number and $\alpha$ is a complex coefficient representing the size and phase of the state.
The effect of loss is the reduction in size of the amplitude and in the limit $|\alpha'|\ll{1}$, we approximate the small amplitude coherent state as $| \alpha' \rangle \approx \mathcal{N} \left( | 0 \rangle + \alpha' | 1 \rangle \right),$
with $\mathcal{N}= \exp \left(- |\alpha'|^2 / 2\right)$.
Here higher order photon number states can be treated as negligibly small.
In this description, the ideal NLA is realised by the state transformation,
\begin{equation} \label{eqn:NLAtruncState}
\mathcal{N} \left(|0\rangle + \alpha'|1\rangle \right) \rightarrow \mathcal{N}'\left(|0\rangle+g\alpha'|1\rangle\right),
\end{equation}
where $\mathcal{N}'=\exp{(-|g\alpha'|^2/2)}$ is the renormalisation after amplification, and $g$ is the amplitude gain.
Thus in the limit of large loss, the small amplitude coherent state can be approximated as a two-level system allowing us to take advantage of photonic qubit logic to construct the weak coupling interaction.

We implement the weak measurement on the signal mode using a controlled-Z (CZ) gate to couple a coherent state in the signal mode to an ancilla single photon in the meter mode.
Formally, the CZ gate is a two-qubit logic gate which performs the Pauli-Z operation on the target qubit conditioned on the input control qubit being in the logical `\textbf{1}' state.
Here we consider the CZ gate configured for polarisation qubits where $|H\rangle$ (horizontal) and $|V\rangle$ (vertical) represent the logical basis states $|\textbf{0}\rangle$ and $|\textbf{1}\rangle$ respectively \cite{Kok2007,RalphPryde2009}.
The gate implements the two-qubit transformation $|V\rangle |V\rangle \rightarrow - |V\rangle |V\rangle$, while remaining basis states $|H\rangle |H\rangle, |H\rangle |V\rangle,$ and $|V\rangle |H\rangle $ undergo identity operations.
Henceforth we adopt the labels of signal ($s$) and meter ($m$) for the control and target modes respectively, as shown in \fref{fig:concept}(b).

In our NLA scheme, the signal mode is occupied by a small coherent state which is not a polarisation qubit but remains operationally compatible with the physical realisation of the gate. This is because putting the vacuum state into the control input of the CZ gate leaves the target unaffected, just as if a logical-zero-encoded polarisation qubit state were used in the control.
Specifically by construction of the CZ gate, we can align the input signal to the vertical polarisation mode of the gate,
\begin{equation} \label{eqn:smlCoherState}
| \psi_{s} \rangle = | \alpha' \rangle_{s,V} \approx \mathcal{N} \left( | 0 \rangle + \alpha' | 1 \rangle \right)_{s,V}.
\end{equation}
As previously stated, the basis states in \eref{eqn:smlCoherState} perform an analogous task to a polarisation qubit, e.g., $|0\rangle_{s,V} \equiv |H\rangle$ and $|1\rangle_{s,V} \equiv |V\rangle$.
Hence small coherent states will coherently interact, as an ideal qubit would, with the meter qubit through the action of the CZ gate which agrees with the result shown in Ref.~\cite{McMahon2014}.

The polarisation qubit in the meter mode is prepared in the form,
\begin{equation} \label{eqn:phi}
| \psi_{m}^{\phi} \rangle = \frac{| H \rangle + i e ^ {i \phi} | V \rangle}{\sqrt{2}},
\end{equation}
where $\phi$ represents the angle of rotation around the Z-axis of the meter state vector in the Bloch (or Poincar\'{e}) sphere representation.
The action of the gate can be summarised as a $\pi$ rotation of the meter vector, around the Z-axis, iff the vertical mode of the signal is occupied by a photon, i.e., $|1\rangle{}_{s,V}$.

Evolution of the signal, \eref{eqn:smlCoherState}, and meter, \eref{eqn:phi}, states through the CZ gate produces the following joint state,
\begin{eqnarray}
| \Psi_{I} \rangle &= \hat{U}_{CZ} | \psi_{s} \rangle \otimes | \psi_{m}^{\phi} \rangle  \nonumber \\
& = \frac{\mathcal{N}}{\sqrt{2}} \left( |0\rangle_{s,V} (|H\rangle + i e^{i\phi} |V\rangle)_m + |1\rangle_{s,V} (|H\rangle - ie^{i\phi} |V\rangle)_m \right). \label{eqn:intState}
\end{eqnarray}
We denote this as the intermediate state $|\Psi_{I}\rangle$, as it precedes the final measurement step, i.e., the projective measurement on the meter.

Our protocol successfully amplifies the signal when the meter qubit is found to be in the state, $| \psi_m^P \rangle = \left(| H \rangle - i | V \rangle \right)_m/\sqrt{2}$.
By post-selecting on this measurement outcome, the signal state $|\psi_s^o\rangle$ is transformed as follows:
\begin{equation} \label{eqn:weakMeasNLA}
|\psi_s^o\rangle = \langle \psi_m^P | \Psi_{I} \rangle = \frac{\mathcal{N}'}{\sqrt{2}} \left(1-e^{i\phi}\right) \left( |0\rangle + g \alpha'|1\rangle\right)_{s,V} .
\end{equation}
Here the amplitude gain, $g = (1+e^{i\phi})/(1-e^{i\phi}),$ depends directly on the measurement strength set by the initial meter state parameter $\phi$. 
Furthermore we calculate the intensity gain,
\begin{equation} \label{eqn:gain_phi}
| g | ^2  = \cot^2 \left( \frac{\phi}{2} \right),
\end{equation}
which tends to infinity as $\phi \rightarrow 0$ . However, the scaling factor (which is related to the probability of success) simultaneously vanishes, i.e., $|\mathcal{N}'(1-e^{i\phi})/\sqrt{2}|^2\rightarrow 0$. 

Thus far we have assumed an ideal construction of a deterministic CZ gate, however it remains non-trivial to attain the necessary non-linearities in optical media to implement it in practice.
Instead we consider a non-deterministic CZ gate, realizable with linear optics and single photon counting, to demonstrate this protocol.
A nondeterministic CZ gate operation (\fref{fig:concept}(b)), conditional on the detection of one and only one photon in the meter output, performs the protocol in the same way as the deterministic gate. However, it introduces a scaling parameter to both the expected intensity gain, i.e.,
\begin{equation} \label{eqn:gain_phi_nondet}
|g'|^2 = \frac{1}{3} \cot^2 \left(\frac{\phi}{2}\right),
\end{equation} 
and the probability of success,
\begin{equation} \label{eqn:Ps_nondet}
P_{s}' = || \langle g'\alpha' | \psi_s^o \rangle ||^2 = \frac{\mathcal{N'}^2}{3} \frac{1}{1+3|g'|^2} (1 + |g'|^2|\alpha'|^2).
\end{equation}

In principle a noiseless linear amplifier is phase insensitive and operates identically for input coherent states $|\alpha'\rangle$ of arbitrary unknown phase.
We can test our device using a coherent state which samples all phases, by considering a phase averaged coherent state \cite{Allevi},
\begin{equation} \label{eqn:PHAV}
\hat{\rho}_{PHAV} = e^{-\frac{|\alpha|^2}{2}} \sum_{n=0}^{\infty} \frac{|\alpha|^n}{\sqrt{n!}} |n\rangle \langle n|.
\end{equation}
Following from previous discussions, we expect for a small amplitude coherent state,
\begin{equation} \label{eqn:PHAV2}
\hat{\rho}_s \approx \mathcal{N} \left( |0\rangle \langle 0| + |\alpha'| |1\rangle \langle 1| \right)_{s,V}.
\end{equation}
We note that the signal is now represented as a diagonalised density operator, i.e., without coherences, however such states exhibit the same photon number statistics as a coherent state.
In the laboratory such a state could be practically prepared by adding loss to mode containing a single photon state.

\section{Experiment and Results \label{sec:experiment}}
\begin{figure}
	\centering
	\includegraphics{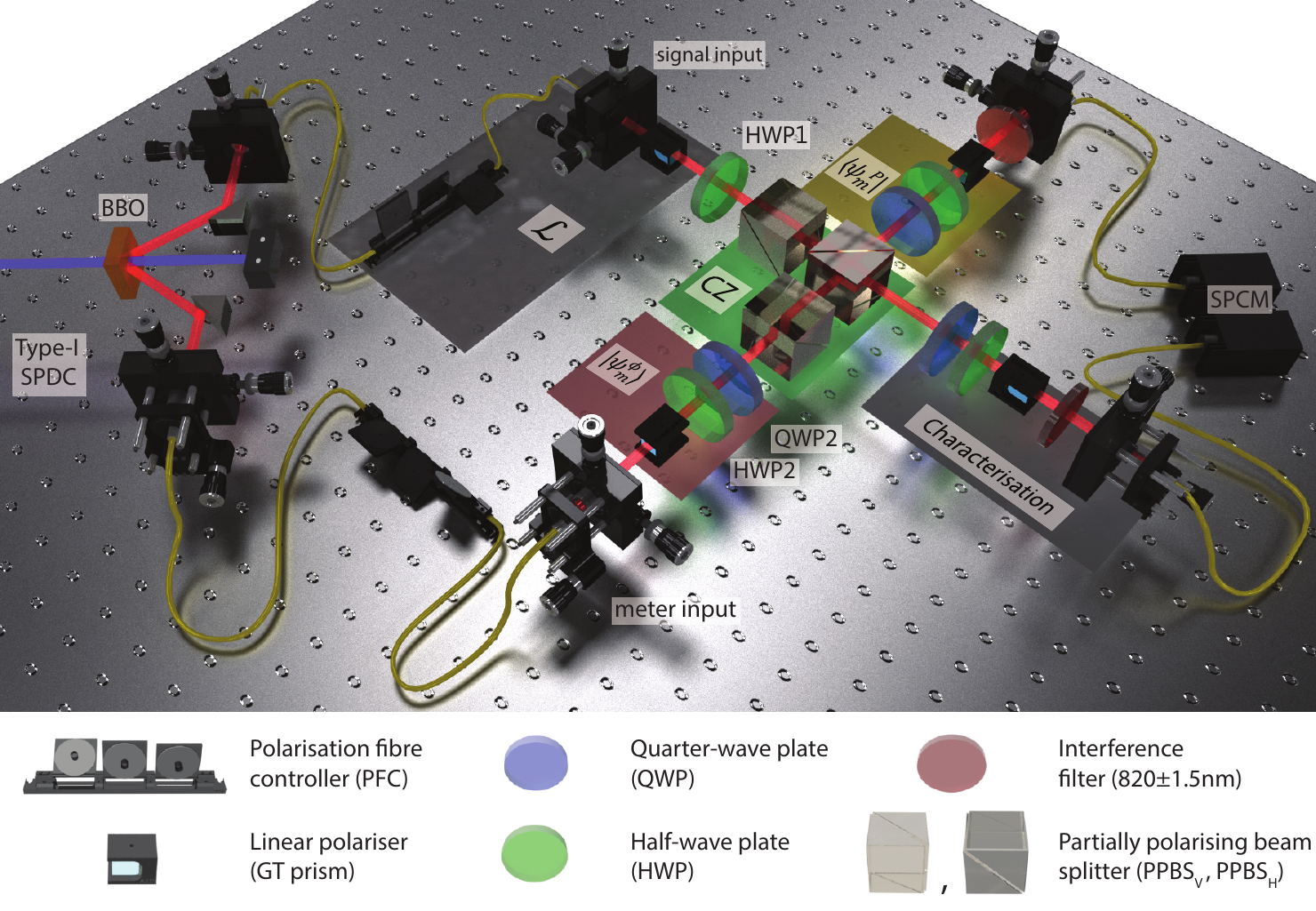}
	\caption{\label{fig:experiment} A BBO crystal (cut for Type-I SPDC) is optically pumped by a 410nm CW diode laser producing photons pairs which are collected at the degenerate condition (e.g., 820nm photons) into single mode fibres.
	Loss, $\mathcal{L}$, is added to the signal mode using a PFC and GT then HWP1 sets the polarisation of the signal state.
	In the meter mode, a GT followed by HWP2 and QWP2 prepares the meter qubit $|\psi_{m}^\phi\rangle$ which sets the measurement strength (and gain).
	The CZ gate is implemented using two PPBS$_H$, one in the signal and meter mode, and one PPBS$_V$ in the centre where the vertical polarised photons interfere non-classically.
	Projective measurements, in the polarisation bases, are implemented on both the signal and meter modes using a QWP, HWP and GT prior to fibre-coupled SPCM.
	}
\end{figure}

We demonstrate the weak-measurement-based amplifier using an optical CZ gate \cite{OBrien2003,Langford2005} as shown in \fref{fig:experiment}.

We produce the single photon states, which we use for both the signal and meter, via spontaneous parametric down conversion (SDPC).
A 2mm thick $\beta$-barium borate (BBO) crystal, phase matched for Type I SPDC, is optically pumped by a 410nm continuous wave (CW) diode laser.
Photon pairs are collected at the degenerate condition (e.g., for 820nm photons) into optical fibres.
To optimise mode matching, which is crucial for implementing the CZ gate, we use single mode fibres and aspheric collimating lenses for all fibre coupling optics which produce near-collimated, Gaussian beam modes in free space at the gate.
Each fibre is mounted in a fibre polarisation controller (FPC), which can be used to undo unwanted unitary rotations arising from fibre birefringence.

The small, phase-averaged coherent state is prepared by adding loss $\mathcal{L}$ to the signal mode at the input to the gate.
The combination of the FPC and linear polariser, i.e., a Glan-Taylor (GT) prism, acts as a variable beam splitter to control the amount of vacuum added to the mode, 
i.e., $\hat{\rho}_s = \mathcal{L} |0\rangle\langle0| + (1-\mathcal{L})|1\rangle\langle1|$.
Using a half-wave plate, HWP1, we align the attenuated signal state to the vertical polarisation mode of the CZ gate.
We also use HWP1 to split part of the signal into the horizontal polarisation mode of the gate, where it serves as a reference arm in an interferometer to investigate the coherence preserving properties.

At the meter input of the gate the transmission is maximised through the GT, producing a high purity polarisation state in $|H\rangle$.
A half- and quarter-wave plate, HWP2 and QWP2 respectively, implement the rotations to prepare the polarisation qubit as \eref{eqn:phi}.
The fibre coupler is mounted on a motorised delay stage to precisely control the temporal mode overlap of the signal and meter modes.

Both the signal and meter mode are coupled to the free space CZ gate which is implemented using three partially polarising beam splitters (PPBS); one aligned to the vertical polarisation (PPBS$_V$), and two in the horizontal polarisation (PPBS$_H$).
With the optic axis of PPBS$_V$ aligned to the vertical polarisation, a 2/3 beam splitter (BS) operation is performed on the vertical components of the signal and meter modes while the orthogonal polarisation (horizontal) is fully transmitted.
Similarly the PPBS$_H$ perform the 2/3 BS operation only on the horizontal mode which balances the probability amplitudes of the polarisation components in the signal and meter.
Using these devices, we implemented the conceptual CZ gate configuration shown in \fref{fig:concept}(b) without spatially separating and recombining the polarisation components of the signal and meter.
We note that in the physical realisation of the gate, the signal and meter modes propagate through the transmitted ports of the 2/3 PPBS instead of the reflected ports of 1/3 BS depicted in the conceptual diagram --- the logical operation is unchanged.

We set up generalised projective measurement stages at both outputs of the CZ gate; these consist of a QWP, HWP and a GT which set the measurement basis.
In the meter arm, this allows us to perform the necessary projective measurement, $\langle\psi_m^p|$ as the final step of the NLA protocol.
Note that the characterisation stage in the signal mode is used to verify the output properties of the amplifier --- its gain and noise characteristics --- and is not required by the protocol.
A narrow band ($\pm{1.5}$nm) filter is placed in front of each fibre-coupled single photon counting module (SPCM) which eliminates spectral distinguishability.

\subsection{Characterising gain performance\label{sec:gain}}
We evaluate the signal gain by comparing the measured state size in the signal mode at the input $|\alpha'|^2$ and output $|g\alpha'|^2$ of the amplifier, from which we obtain the intensity gain, i.e., $|g\alpha'|^2/|\alpha'|^2 = |g|^2$.
We note that the magnitude $|\alpha|^2$ for small amplitude phase averaged coherent states is directly related to the probability weighting of the single photon state $|\mathcal{N}\alpha|^2$.
By working with small input states, we find the normalisation constant, $\mathcal{N}^2$, remains close to unity which allows us to directly estimate the state size by determining the probability of measuring a single photon in the signal mode.
Similarly, if the amplified state at the output remains small, we can measure the state size in the same way.

At the output of the amplifier we are interested only in the cases when the state has successfully amplified, i.e., those heralded by the detection of a photon in the meter mode following the projective measurement $\langle \psi_m^P|$.
Thus we measure the photon detection events in coincidence.
To determine the conditional probability of measuring a single photon in the signal mode, we divide the coincidence detections in both modes by the single detection events in the meter mode.
We use this method also to evaluate the single photon probability at the input mode to ensure a fair comparison of the measured quantities.

\begin{figure}
	\includegraphics{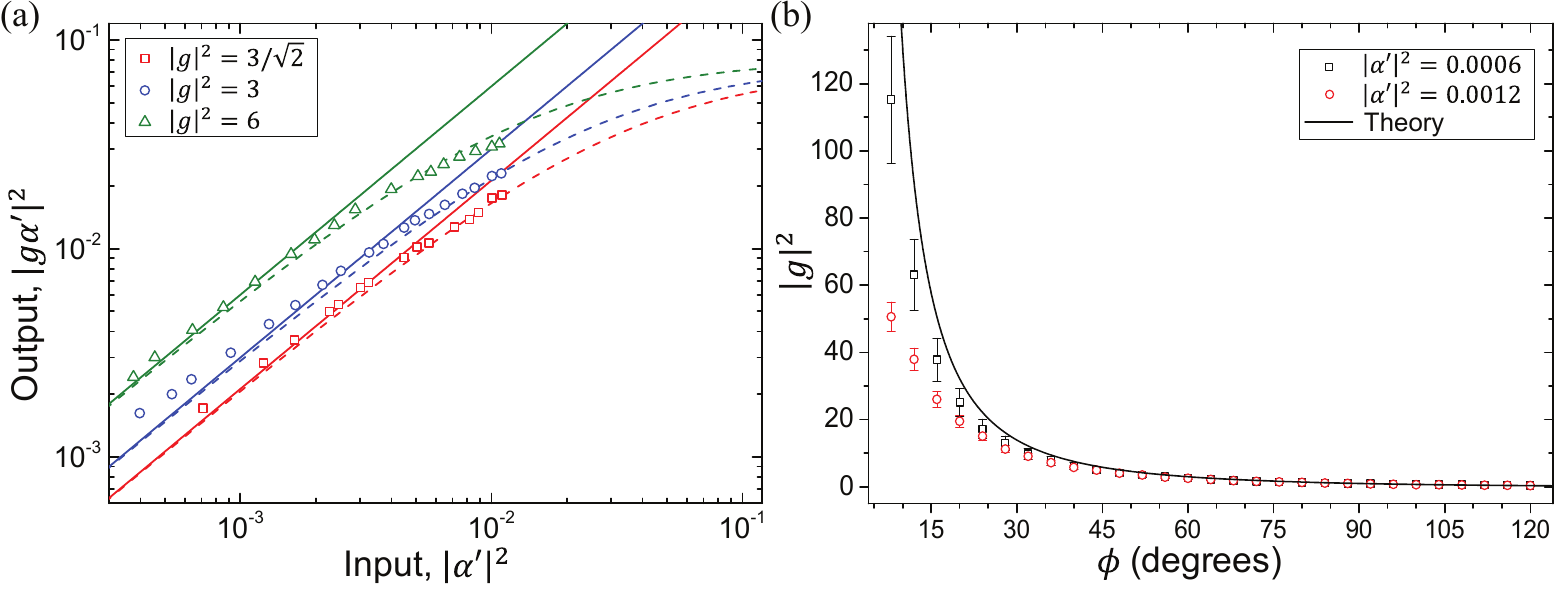}
	\caption{\label{fig:gain} (a) The input and output state size for nominal gain settings of the amplifier $|g|^2=3/\sqrt{2}$, 3, and 6. Theoretical curves are plotted assuming ideal conditions (solid line) and modelled with source inefficiencies (dashed line). (b) For two nominal input state sizes $|\alpha'|^2 = 0.0006$ (black squares) and $|\alpha'|^2=0.0012$ (red circles) the gain was examined as function of the measurement strength, $\phi$. The ideal theoretical gain is shown (solid line). In all cases error bars have been calculated using standard error propagation and assuming Poissonian statistics.}
\end{figure}

In the experiment the detectors remain positioned at the end of the setup thus all measurements on the signal and meter modes are made after the gate.
To measure the input signal state we prepare the meter qubit in $|H\rangle$ and change the projective measurement to $\langle H|$ this ensures that no interaction, and thus no amplification, occurs through the gate.
By performing all measurements after the gate, a raw measurement of the input state size acquires a $1/3$ scaling factor owing to the PPBSs, i.e., $|\alpha'|^2 = 1/3 |\alpha'|^2_{act}$.
To correctly analyse our raw state size measurements, we correct the theoretically expected gain by using \eref{eqn:gain_phi} which includes this extra scaling factor of 3.

We first investigate the response of the amplifier as a function of the input state size, $|\alpha'|^2$.
We prepare the meter qubit using \eref{eqn:gain_phi} to set $\phi$ for three nominal gain settings; $|g|^2 = 3/\sqrt{2}$, 3, and 6, and measure the output state size over a range of input sizes.
The experimental results are shown in \fref{fig:gain}(a), along with the expected gain plotted as solid theoretical curves which assumes perfect state preparation of the meter qubit in an ideal non-deterministic CZ gate.
We observe linear gain in the regime of small input state sizes, agreeing with the theory, however as the state size increases the output state sizes begins to deviate.
This is expected in part due to renormalisation that takes place as the state size becomes larger, and thus the truncated coherent state becomes a less accurate approximation.
However, at the state sizes observed, this does not fully explain the results.
One might suspect that the source inefficiencies, specifically the limited heralding efficiency set by the non-collinear geometry of the SPDC source, play a larger role in the observed discrepancy.
The heralding efficiency defines the conditional probability of detecting a photon from one arm of the source provided a detection event is present in the other arm.
In our experimental setup, this bounds the maximally attainable state size which is defined by a similar conditional probability, i.e., the heralding mechanism which flags the successful amplification.
As such, an additional renormalisation factor is needed to scale the actual attainable states observed in the signal mode.
Using the method described in Ref.~\cite{Xiang2010}, we theoretically modelled the expected output state size (dashed lines) accounting for an estimated source heralding efficiency of $\epsilon = 0.35$, which omits detector efficiencies.

The ideal solid curves assume perfect state preparation of the meter qubit, however manufacturing imperfections in the optics will introduce an unwanted phase, and rotate the meter state vector.
We expect this to be the cause of the larger-than-expected output state sizes observed when we should have ideally prepared the amplifier to implement a gain of $|g|^2=3$.
In \fref{fig:gain}(a), we have calculated uncertainties assuming Poissonian statistics from our photon detection schemes and carried out error propagation, though we find in all cases the error bars were smaller than data points recorded.

We also investigate the gain of the amplifier as a function of $\phi$, i.e., the measurement strength, for two nominal input state sizes, $|\alpha'|^2 = 0.0006$ and $|\alpha'|^2 =0.0012$ as shown in \fref{fig:gain}(b).
We observe that, for small nominal gain values, the experimental gain agrees well with theory.
As we operate the amplifier in the high gain regimes the output state size, and thus the gain we achieve, begins to deviate from theory owing to the limitations in the source efficiency discussed previously.
In particular, we note that the larger input, i.e., $|\alpha'|^2=0.0012$, deviates more rapidly from theory and reaches a lower maximal gain; this is because the net output state size grows faster with a larger input and reaches the saturating condition more rapidly.

Finally we note that as the gain is increased, i.e., $\phi\rightarrow0$, the error bars increase also. 
This is due to the diminishing probability of successfully amplifying the state with larger gain, as described by \eref{eqn:Ps_nondet}, leading to less coincidence (or heralded) events being measured.
Naturally for the smaller input state, $|\alpha|^2=0.0006$, the number of coincidences observed within the measurement integration time is decreased giving rise to larger uncertainties.

\subsection{Evaluating noise properties\label{sec:noise}}
In operating an amplifier it is important to have knowledge of any noise added at the output.
To test this, we examine the coherence-preserving property of the amplifier by integrating our amplifier in an interferometer and measuring the fringe visibility.
As shown in \fref{fig:concept}(b), the amplifier only acts on the signal in the vertical polarisation mode which leaves the horizontal polarisation mode free as the reference arm for our interferometer.
To account for the role of the amplifier acting only on one arm, we bias the interferometer by setting the ratio of $H:V$ in the signal mode to $|g|^2:1$, using HWP1, such that following the amplification of the vertical mode the two arms interfere optimally in principle.
Any added noise in the amplified mode will therefore reduce the measured fringe visibility.

\begin{figure}
	\centering 
	\includegraphics{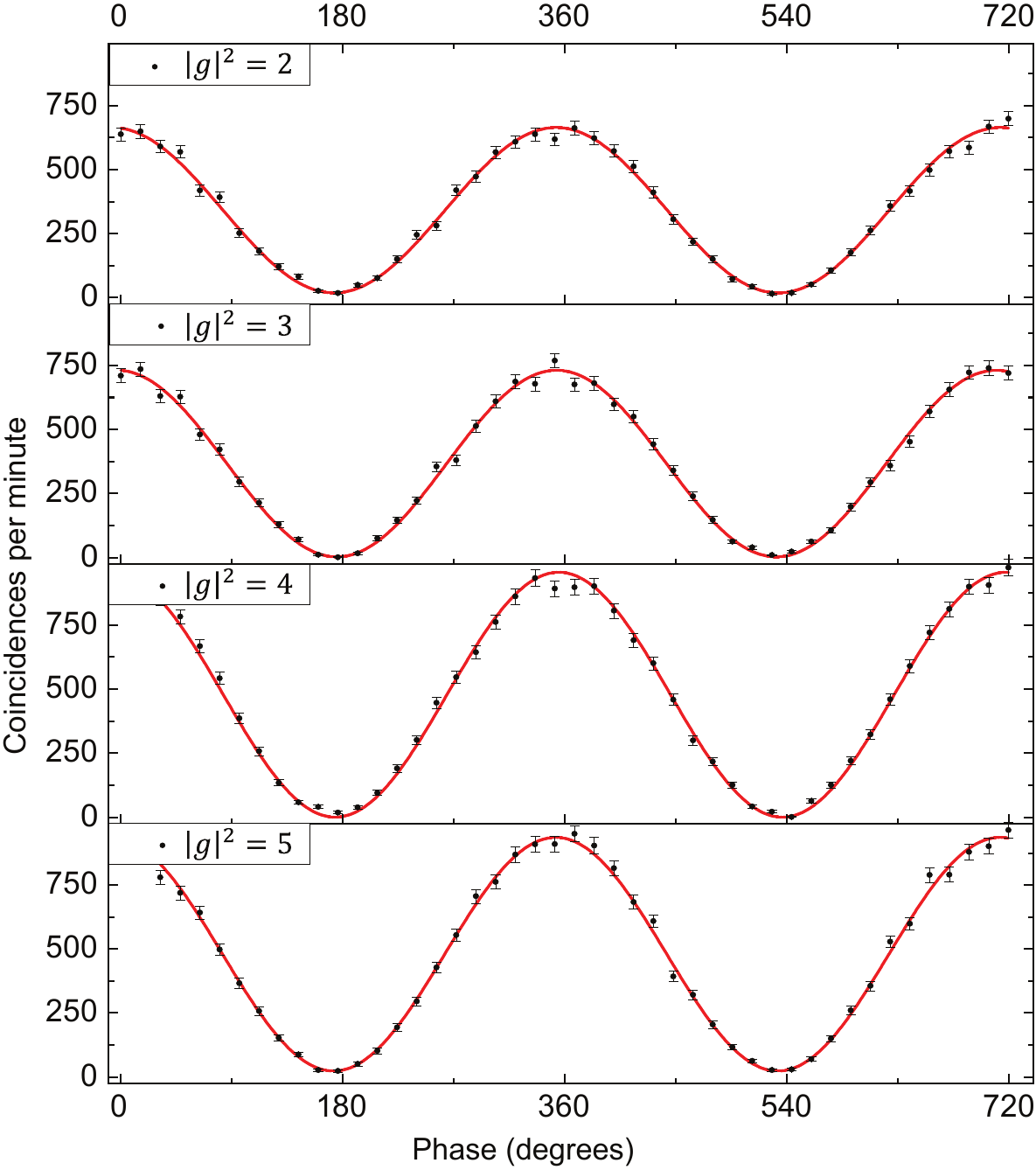}
	\caption{\label{fig:noise} High visibility interference fringes, for four gain settings $|g|^2 = 2,3,4,$ and 5, indicates coherence in the signal mode is preserved after the amplification.
	The calculated visibilities from the fitted datasets were $0.94\pm{0.02}, 0.99\pm{0.02}, 1.00\pm{0.02}$, and $0.95\pm{0.02}$ respectively.
	Measured data points are plotted with error bars showing one standard deviation, assuming Poissonian statistics.}
\end{figure}

To measure the interference fringes in the interferometer, we use the characterisation stage in the signal output to coherently combine the two polarisation modes which form the arms of the interferometer and sample the relative phase between them.
By measuring in coincidence with the heralding condition in the meter mode, we only consider cases where the vertical mode is successfully amplified.
In light of the results from the gain measurements, we prepared a very small input state size $|\alpha'|=0.0015$ in the vertical polarisation mode to limit the effects of saturation at the output.
Furthermore, from the results in \fref{fig:gain}(b), we use the experimentally measured gain values to calibrate the meter qubit preparation to optimise the gain settings.
We performed the fringe measurement using four gain values ($|g|^2=2,3,4,$ and 5) as shown in \fref{fig:noise}.
The experimental data points are fitted assuming fringes of the form $\cos(\varphi)$, and the visibility in all four cases was found to be well above $0.90$ which indicates a high level of coherence in the output states of the interferometer.

For the four nominal gain settings we calculate the theoretical visibility attainable if a linear, phase-preserving amplifier is used \cite{RalphPC}; 0.71, 0.58, 0.50, and 0.45 respectively.
We find that our measured visibilities are significantly larger than these classical limits, as expected for noiseless operation of the amplifier \cite{Xiang2010}.

\section{Conclusion \label{sec:conclusion}}
We have experimentally demonstrated that the measurement back-action of a weak measurement can be used to amplify the quantum state of a single harmonic oscillator---here, an optical mode. The weak measurement is implemented by coupling the system to an ancilla, the meter, using a quantum logic gate.
Following a projective measurement of the ancilla system, the signal mode is amplified conditional on obtaining the correct state in the meter mode.
Our experimental results show for small input state sizes, the output states follow a linear gain as expected.
Owing to the moderate heralding efficiency of the source, the range of input sizes which follow the linear amplification was restricted. However, this is not a fundamental restriction, as high-heralding-efficiency sources are starting to become available \cite{Weston2016}.
Furthermore, amplifiers for small quantum states can form component stages in a larger amplifier architecture \cite{Xiang2010,Kocsis2012}
We find that in our experiment we achieved large heralded intensity gains. The amplified states were found to preserve coherences in the signal optical mode which would not be possible if a noisy, linear phase-preserving amplifier was used instead. In principle, a weak measurement based approach can provide optimal performance in the low gain limit \cite{McMahon2014} and provides a general framework for realising heralded amplifiers as deterministic entangling operations for optical systems begin to be realised. 

\section*{Acknowledgements}
This work was performed by the Australian Research Council Centre of Excellence for Quantum Computation and Communication Technology (CE110001027). We thank Tim Ralph and Raj Patel for helpful discussions.

\end{document}